\documentclass[prd,amsmath,amssymb,amsthm,superscriptaddress,nofootinbib]{revtex4}

\allowdisplaybreaks
\usepackage[normalem]{ulem}
\usepackage{xcolor}
\usepackage[mathscr]{euscript}
\usepackage{bm}
\usepackage{graphicx}
\usepackage{subfigure}
\usepackage[colorlinks=true,linkcolor=red]{hyperref}

\newcommand{\bea}{\begin{eqnarray}}
\newcommand{\eea}{\end{eqnarray}}
\newcommand{\beq}{\begin{equation}}
\newcommand{\eeq}{\end{equation}}
\newcommand{\nn}{\nonumber}
\def\/{\over}

\begin{document}

\title{Entanglement dynamics for atoms near a reflecting boundary: Enhancement and suppression by environment-induced interactions }

\author{Ying Chen}
\affiliation{School of Physics and Astronomy, China West Normal University, Nanchong, Sichuan 637002, People’s Republic of China}
\author{Hongwei Yu}
\email[Corresponding author. ]{hwyu@hunnu.edu.cn}
\affiliation{Department of Physics, Key Laboratory of Low Dimensional Quantum Structures and Quantum Control of Ministry of Education, and Hunan Research Center of the Basic Discipline for Quantum Effects and Quantum Technologies, Hunan Normal University, Changsha, Hunan 410081, China}
\author{Jiawei Hu}
\email[Corresponding author. ]{jwhu@hunnu.edu.cn}
\affiliation{Department of Physics, Key Laboratory of Low Dimensional Quantum Structures and Quantum Control of Ministry of Education, and Hunan Research Center of the Basic Discipline for Quantum Effects and Quantum Technologies, Hunan Normal University, Changsha, Hunan 410081, China}

\begin{abstract}

We investigate how environment-induced interactions influence the entanglement dynamics of two atoms held at fixed positions near a perfectly reflecting boundary. Within the framework of open quantum systems, we explicitly incorporate the environment-induced energy shifts, including both atom-boundary contributions and an environment-induced atom-atom interaction, which are often neglected in previous studies.  We show that, for any initial two-atom state, these energy-shift effects qualitatively and quantitatively modify the entanglement dynamics relative to treatments that omit them. Depending on the geometry and parameter regime, the environment-induced interactions can either enhance entanglement generation---yielding a larger maximum concurrence and a longer entanglement lifetime---or suppress it, reducing both the peak concurrence and the survival time. This behavior contrasts sharply with the free-space case, where the environment-induced atom-atom interaction affects entanglement generation only for a restricted class of initial states and does so in an exclusively assisting manner.

\end{abstract}

\maketitle

\section{Introduction}
Quantum entanglement is among the most distinctive features of quantum physics and underpins key quantum technologies, such as quantum communication \cite{CH1} and quantum teleportation \cite{CH2}. In realistic settings, however, any quantum system inevitably interacts with its surrounding environment. Even when the environment is the vacuum, the system remains subjected to vacuum fluctuations.
Such coupling leads to decoherence and entanglement degradation. For example, two initially entangled atoms in vacuum may become completely disentangled within a finite time, a phenomenon known as entanglement sudden death \cite{Yu1,Yu2}. 

The environment can also play a constructive role. In particular, field-mediated correlations may induce effective interactions between subsystems that do not directly couple, thereby enabling entanglement creation under suitable conditions \cite{Braun,Kim,ss,Basharov,Jakobczyk,Benatti1,Benatti2,br1,Ficek4,Ficek5,Ficek3,Tana}.
For two atoms immersed in a thermal bath with finite separation, entanglement generation occurs only when the interatomic separation and the bath temperature meet specific conditions \cite{Tana,Benatti2}, while entanglement sudden death is a generic  feature \cite{Tana}. Moreover, entanglement can be recreated after disappearing, a phenomenon referred to as entanglement revival \cite{zf}.

In many such studies, the correlations of fluctuating quantum fields are central to entanglement dynamics. When the spacetime (or effective environment) has a nontrivial topology, vacuum field fluctuations are accordingly modified, giving rise to novel features in entanglement dynamics. One of the simplest and  most instructive examples  is the presence of a reflecting boundary. The effects of a boundary on the entanglement dynamics of a detector-field system have been investigated in Ref. \cite{Hu-2013}, where the behavior of detector-field entanglement can be qualitatively explained using the method of mirror images. The conditions for early-time entanglement generation and the full entanglement evolution of a two-atom system near a boundary have been investigated  in Refs. \cite{Zhang1,Zhang2} and \cite{Cheng}, respectively, which show that the boundary can offer additional handles for  manipulating the entanglement dynamics.  Related effects have also been explored in other topologically nontrivial settings, such as cosmic string spacetimes \cite{He} and  a cylindrical universe \cite{LB,Ed}.

From the Gorini-Kossakowski-Lindblad-Sudarshan (GKLS) master equation \cite{Kossakowski,Lindblad,Breure} describing the dynamics of an open quantum system, one sees that the environment induces  decoherence and dissipation on the one hand and an energy shift on the other.  For a two-atom system, the energy shift term contains  both  individual Lamb shifts and an environment-induced interatomic interaction. In a recent work  \cite{Chen}, the influence of the environment-induced interatomic interaction on the entanglement dynamics of two uniformly accelerated atoms in the Minkowski vacuum was investigated. Compared with earlier treatments neglecting this interaction \cite{Benatti3,Lima20,Hu}, entanglement generation can be enhanced for certain initial states, while the anti-Unruh behavior in terms of the entanglement generation disappears.

The presence of a boundary makes the situation even more intriguing: the environment-induced Lamb shifts become position dependent, so the environment-induced interaction includes both an atom-boundary contribution and an induced atom-atom interaction. Previous studies of two-atom entanglement dynamics near a boundary~\cite{Zhang1,Zhang2,Cheng} have primarily focused on decoherence and dissipation, typically neglecting the environment-induced energy shifts. It is therefore important to clarify how these energy-shift effects modify the entanglement dynamics and whether they can enhance or suppress entanglement generation. This is the goal of the present work.   Throughout, we employ natural units with  $\hbar=c=1$, where $c$ is the speed of light, and $\hbar$ is  the reduced Planck constant.

\section{The Master Equation}
We consider an open quantum system composed of two atoms weakly coupled to a fluctuating massless scalar field in the Minkowski vacuum in the presence of a perfectly reflecting boundary. The total Hamiltonian of the system is
  \begin{equation}\label{pf1}
H=H_{S}+H_{F}+H_{I}.
\end{equation}
Here
\begin{equation}\label{hs}
H_{S}=\frac{\omega}{2}\sigma_{3}^{(1)}+\frac{\omega}{2}\sigma_{3}^{(2)}
\end{equation}
is the Hamiltonian of the two-atom system, where $\sigma^{(1)}_{i}=\sigma_{i}\otimes\sigma_{0}$ and $\sigma^{(2)}_{i}=\sigma_{0}
\otimes\sigma_{i} $, with $\sigma_{i}~(i=1,2,3)$ being the Pauli matrices, $\sigma_{0}$ the $2\times2$ unit matrix, and $\omega$  the energy level spacing of the atoms.  $H_{F}$ denotes the free Hamiltonian of the fluctuating scalar field;  its explicit form is not needed here. We take the interaction Hamiltonian $H_{I}$  in a form analogous to electric-dipole coupling \cite{Audretsch1994}
\beq
H_{I}=\mu[\sigma^{(1)}_{2}\Phi(t,x_{1})+\sigma^{(2)}_{2}\Phi(t,x_{2})],
\eeq
where $\mu$ is the coupling constant, and $\Phi(t,x)$ denotes a fluctuating massless scalar field in the Minkowski vacuum.

We assume an initially factorized state  $ \rho_{\rm tot}(0)=\rho(0)\otimes|0\rangle\langle0|$, where $\rho(0)$  is the initial two-atom state, and $|0\rangle$ is the vacuum state of the field. In the weak-coupling limit,  the reduced dynamics of the two-atom system obeys a GKLS master equation \cite{Kossakowski,Lindblad,Breure}
\beq\label{master1}
\frac{\partial\rho(\tau)}{\partial\tau}=-i[H_{\rm eff},\rho(\tau)]+
\mathcal{D}[\rho(\tau)],
\eeq
with
\beq\label{master2}
H_{\rm eff}=H_{S}-\frac{i}{2}\sum^{2}_{\alpha,\beta=1}\sum^{3}_{i,j=1}
H_{ij}^{(\alpha\beta)}\sigma_{i}^{(\alpha)}\sigma_{j}^{(\beta)}
\eeq
and
\beq\label{master3}
\mathcal{D}[\rho(\tau)]=\frac{1}{2}\sum^{2}_{\alpha,\beta=1}\sum^{3}_{i,j=1}
C_{ij}^{(\alpha\beta)}[2\sigma_{j}^{(\beta)}\rho\sigma_{i}^{(\alpha)}-
\sigma_{i}^{(\alpha)}\rho\sigma_{j}^{(\beta)}-\rho\sigma_{i}^{(\alpha)}
\sigma_{j}^{(\beta)}].
\eeq
Here the coefficients of the Kossakowski matrix $C_{ij}^{(\alpha\beta)}$
and  the effective Hamiltonian $H_{\rm eff}$ are determined by the Fourier transform of the field correlation function $\langle\Phi(\tau,x_{\alpha})\Phi(\tau',x_{\beta})\rangle$,
\beq
\mathcal{G}^{(\alpha\beta)}(\lambda)=\int^{\infty}_{-\infty}d\Delta\tau
e^{i\lambda\Delta\tau}\langle\Phi(\tau,x_{\alpha})\Phi(\tau',x_{\beta})\rangle,
\eeq
and its Hilbert transform,
\beq
\mathcal{K}^{(\alpha\beta)}(\lambda)=\frac{P}{\pi i}\int^{\infty}_{-\infty}d\omega
\frac{\mathcal{G}^{(\alpha\beta)}(\omega)}{\omega-\lambda},
\eeq
with $P$ representing the principal value. Then, the coefficient matrix $C_{ij}^{(\alpha\beta)}$ can  be written as
\beq \label{xi1}
C_{ij}^{(\alpha\beta)}=A^{(\alpha\beta)}\delta_{ij}-iB^{(\alpha\beta)}
\epsilon_{ijk}\delta_{3k}-A^{(\alpha\beta)}\delta_{3i}\delta_{3j},
\eeq
where
\bea
\nn A^{(\alpha\beta)}=\frac{\mu^{2}}{4}[\mathcal{G}^{(\alpha\beta)}(\omega)
+\mathcal{G}^{(\alpha\beta)}(-\omega)],\\
B^{(\alpha\beta)}=\frac{\mu^{2}}{4}[\mathcal{G}^{(\alpha\beta)}(\omega)
-\mathcal{G}^{(\alpha\beta)}(-\omega)].
\eea
Similarly,  $H_{ij}^{(\alpha\beta)}$ is obtained by replacing $\mathcal{G}^{(\alpha\beta)}$ with $\mathcal{K}^{(\alpha\beta)}$ in the  expressions above.

The effective Hamiltonian can be decomposed as $H_{\rm eff}=\tilde{H_{s}}+H^{(12)}_{\rm eff}$.  The first part is  a renormalization of the atomic Hamiltonian, which takes the same form as the Hamiltonian (\ref{hs}), but with a redefined energy-level spacing,
\bea
\tilde{\omega}_{\alpha}=\omega-\frac{i\mu^{2}}{2}[\mathcal{K}^{(\alpha\alpha)}(\omega)-
\mathcal{K}^{(\alpha\alpha)}(-\omega)]\;\;\;\;\;\;\;\;\;\;\; (\alpha=1,2).
\eea
In the presence of a boundary, this renormalized energy spacing becomes position dependent. We refer to this position-dependent shift as the environment-induced atom–boundary interaction. 
The second part denotes an environment-induced  coupling between the two atoms
\bea\label{eq12}
H^{(12)}_{\rm eff}=-\sum_{i,j=1}^{3}\Omega_{ij}^{(12)}(\sigma_{i}\otimes\sigma_{j}),
\eea
where
\bea \label{xi2}
\Omega_{ij}^{(12)}=\frac{i\mu^{2}}{4}\{[\mathcal{K}^{(12)}(\omega)+\mathcal{K}^{(12)}(-\omega)]\delta_{ij}-
[\mathcal{K}^{(12)}(\omega)+\mathcal{K}^{(12)}(-\omega)]\delta_{3i}\delta_{3j}\}.
\eea
Finally, the master equation can be written explicitly as
\bea
\nn \frac{\partial\rho(\tau)}{\partial\tau}=&-&i\sum_{\alpha=1}^{2}
{\tilde{\omega}_{\alpha}}[\sigma_{3}^{(\alpha)},\rho(\tau)]+i
\sum_{i,j=1}^{3}\Omega_{ij}^{(12)}[\sigma_{i}\otimes\sigma_{j},\rho(\tau)]\\
&+&\frac{1}{2}\sum^{2}_{\alpha,\beta=1}\sum^{3}_{i,j=1}
C_{ij}^{(\alpha\beta)}[2\sigma_{j}^{(\beta)}\rho\sigma_{i}^{(\alpha)}-
\sigma_{i}^{(\alpha)}\rho\sigma_{j}^{(\beta)}-\rho\sigma_{i}^{(\alpha)}
\sigma_{j}^{(\beta)}].
\eea

\section{Entanglement dynamics for a two-atom system near a boundary}

\begin{figure}[htbp]
  \centering
   \includegraphics[width=0.5\textwidth]{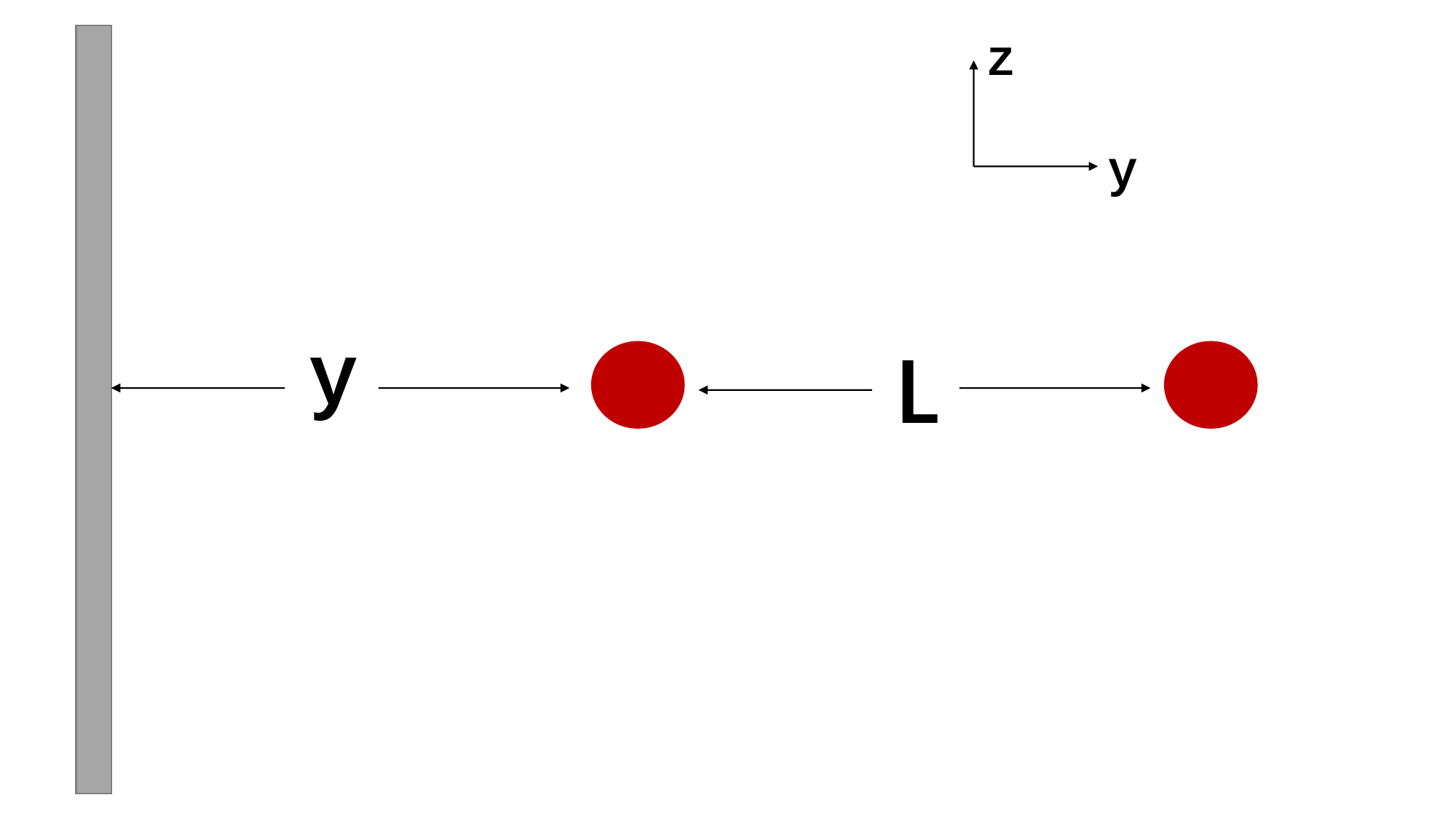}\;\;
  \caption{Two atoms separated by a distance $L$, aligned vertically to the reflecting boundary.
}\label{tu1}
\end{figure}

We now investigate the entanglement dynamics of a two-atom system near a reflecting boundary,  focusing  on how entanglement generation is affected by environment-induced energy shifts. In the presence of a boundary, the environment-induced energy shifts of individual atoms become position dependent; accordingly, the environment-induced interaction comprises both atom-boundary contributions and the induced atom-atom interaction.  We are therefore particularly interested in the situation where the energy shifts of the two atoms differ. So, we place the atoms along a line perpendicular to the boundary, with worldlines
\bea\label{traj15}
\nn &&t_{1}(\tau)=\tau,\;\;x_{1}(\tau)=0,
\;\;y_{1}(\tau)=y,\;\;z_{1}(\tau)=0,\\
&&t_{2}(\tau)=\tau,\;\;x_{2}(\tau)=0,
\;\;y_{2}(\tau)=y+L,\;\;z_{2}(\tau)=0,
\eea
as shown in Fig. \ref{tu1}. Here, $\tau$ is the time coordinate, $y$ is the distance from the boundary to the  nearer atom,  and $L$ is the distance between the two atoms.

The reflecting boundary is located at $y=0$. The field correlation function can be obtained using the method of images, which takes the following form,
\bea
\langle\Phi(t,x_{\alpha})\Phi(t',x_{\beta})\rangle=-\frac{1}{4\pi^{2}}
\frac{1}{(t-t'-i\epsilon)^{2}-(x-x')^{2}-(y-y')^{2}-(z-z')^{2}}\\ \nn
+\frac{1}{4\pi^{2}}\frac{1}{(t-t'-i\epsilon)^{2}-(x-x')^{2}-(y+y')^{2}-(z-z')^{2}},
\eea
where $\epsilon$ is a positive infinitesimal. The Fourier transforms of the correlation functions are
\bea
\mathcal{G}^{(11)}(\lambda)&=&\frac{\lambda}{2\pi}[1-f_{1}(\lambda,  y)]\theta(\lambda),\\
\mathcal{G}^{(22)}(\lambda)&=&\frac{\lambda}{2\pi}[1-f_{1}(\lambda, y+L)]\theta(\lambda),\\
\mathcal{G}^{(12)}(\lambda)&=&\mathcal{G}^{(21)}(\lambda)=\frac{\lambda}{2\pi}[f_{1}(\lambda, L/2)-f_{1}(\lambda, y+L/2)]\theta(\lambda),
\eea
where $\theta(\lambda)$ is the Heaviside step function, which is 0 when $\lambda<0$, and 1 when $\lambda\geq0$, and
\bea
f_{1}(\omega, y)=\frac{\sin(2\omega y)}{2\omega y}.
\eea
Then, according to Eqs. (\ref{xi1}) and (\ref{xi2}), the coefficient matrices $C_{ij}^{(\alpha\beta)}$ and $\Omega^{(12)}_{ij}$ can be written as,
\bea\label{cij19}
&&C_{ij}^{(11)}=B_{1}\delta_{ij}-i B_{1}\epsilon_{ijk}\delta_{3k}-B_{1}\delta_{3i}\delta_{3j},\\
&&C_{ij}^{(22)}=B_{2}\delta_{ij}-i B_{2}\epsilon_{ijk}\delta_{3k}-B_{2}\delta_{3i}\delta_{3j},\\
&&C_{ij}^{(12)}=C_{ij}^{(21)}=B_{3}\delta_{ij}-i B_{3}\epsilon_{ijk}\delta_{3k}-B_{3}\delta_{3i}\delta_{3j},\\
&&\Omega^{(12)}_{ij}=D\delta_{ij}-D\delta_{3i}\delta_{3j},\label{eq24}
\eea
where
\bea\label{b1}
 B_{1}&=&\frac{\Gamma_{0}}{4}[1-f_{1}(\omega, y)],\\
 B_{2}&=&\frac{\Gamma_{0}}{4}[1-f_{1}(\omega, y+L)],\\
 B_{3}&=&\frac{\Gamma_{0}}{4}[f_{1}(\omega, L/2)-f_{1}(\omega, y+L/2)],\\
 D&=&\frac{\Gamma_{0}}{4}[f_{2}(\omega, L/2)-f_{2}(\omega, y+L/2)],\label{d}
\eea
with $\Gamma_{0}=\frac{\mu^{2}\omega}{2\pi}$ being the spontaneous emission rate of inertial atoms in free space and
\bea\label{f2}
f_{2}(\omega, y)=\frac{\cos(2\omega y)}{2\omega y}.
\eea

In order to describe the dynamics of the system, we work in the coupled basis $\{|G\rangle=|00\rangle,|A\rangle=\frac{1}{\sqrt{2}}(|10\rangle-|01\rangle),
|S\rangle=\frac{1}{\sqrt{2}}(|10\rangle+|01\rangle),|E\rangle=|11\rangle\}$.
In this basis, a set of equations describing the time evolution of the density matrix elements can be obtained as
\bea\label{zhufangcheng1}
\nn \dot{\rho}_{GG}&=&2(B_{1}+B_{2}-2B_{3})\rho_{AA}+2(B_{1}+B_{2}+2B_{3})\rho_{SS}+2(B_{1}-B_{2})(\rho_{AS}+\rho_{SA}),\\
\nn \dot{\rho}_{EE}&=&-4(B_{1}+B_{2})\rho_{EE},\\
\nn\dot{\rho}_{AA}&=&-2(B_{1}+B_{2}-2B_{3})\rho_{AA}+2(B_{1}+B_{2}-2B_{3})\rho_{EE}+(-B_{1}+B_{2})(\rho_{AS}+\rho_{SA})-i\Delta(\rho_{AS}-\rho_{SA}),\\
 \nn\dot{\rho}_{SS}&=&-2(B_{1}+B_{2}+2B_{3})\rho_{SS}+2(B_{1}+B_{2}+2B_{3})\rho_{EE}+(-B_{1}+B_{2})(\rho_{AS}+\rho_{SA})+i\Delta(\rho_{AS}-\rho_{SA}),\\
\nn \dot{\rho}_{AS}&=&-2(B_{1}+B_{2}+2iD)\rho_{AS}+2(-B_{1}+B_{2})\rho_{EE}+(-B_{1}+B_{2})(\rho_{SS}+\rho_{AA})+i\Delta(\rho_{SS}-\rho_{AA}),\\
\nn \dot{\rho}_{SA}&=&-2(B_{1}+B_{2}-2iD)\rho_{SA}+2(-B_{1}+B_{2})\rho_{EE}+(-B_{1}+B_{2})(\rho_{SS}+\rho_{AA})-i\Delta(\rho_{SS}-\rho_{AA}),\\
\dot{\rho}_{GE}&=&-2(B_{1}+B_{2})\rho_{GE},\;\;\;\;\;\;\;\;\;\;\;\;\;\;\;\;\;\;\;\;\;
\;\;\;\;\;\;\;\;\;\;\;\;\;\;\;\;\;\;\;\;\;\;\;\;\;\dot{\rho}_{EG}=-2(B_{1}+B_{2})\rho_{EG},\label{zhufangcheng2}
\eea
where $\rho_{IJ}=\langle I|\rho|J\rangle,I,J\in\{G,E,A,S\}$, and
\beq\label{delta}
\Delta=\tilde{\omega}_{2}-\tilde{\omega}_{1}=\frac{\Gamma_{0}}{4}[f_{2}(\omega, y+L)-f_{2}(\omega, y)]
\eeq
is the difference in the environment-induced energy shifts of the two atoms. 
In this paper, we focus on how entanglement generation is affected by environment-induced interactions. Here, $\Delta$ denotes the difference between the environment-induced atom-boundary interactions. The coefficient $D$, which is related to $\Omega_{ij}$ as shown in Eq.~\eqref{eq24}, characterizes the environment-induced atom-atom interaction, since $\Omega_{ij}$ describes the effective coupling between the two atoms as shown in Eq.~\eqref{eq12}. 
For simplicity,  we choose an initial X state, i.e., a density matrix whose only nonvanishing elements are those on the main diagonal and the antidiagonal (the secondary diagonal running from the top-right to the bottom-left corner). In the basis $\{|G\rangle,|A\rangle,|S\rangle,|E\rangle\}$, these correspond to $\rho_{EE},\rho_{AA},\rho_{SS},\rho_{GG}$ and $\rho_{EG},\rho_{GE},\rho_{AS},\rho_{SA}$. 
Equation~(\ref{zhufangcheng2}) shows that these elements evolve independently of the remaining components, so the X structure is preserved.

We quantify entanglement using concurrence \cite{Wootters}. For X states, the concurrence can be calculated as,
\bea
C[\rho(\tau)]=\max\{0,K_{1}(\tau),K_{2}(\tau)\},
\eea
where
\bea
\nn K_{1}(\tau)&=&\sqrt{[\rho_{AA}(\tau)-\rho_{SS}(\tau)]^{2}-[\rho_{AS}(\tau)-\rho_{SA}(\tau)]^{2}}
-2\sqrt{\rho_{GG}(\tau)\rho_{EE}(\tau)},\\
K_{2}(\tau)&=&2|\rho_{GE}(\tau)|-\sqrt{[\rho_{AA}(\tau)+\rho_{SS}(\tau)]^{2}-
[\rho_{AS}(\tau)+\rho_{SA}(\tau)]^{2}}.\label{k2}
\eea

Before proceeding to explicit discussions, we emphasize a key distinction from the boundary-free case. 
 From Eqs. (\ref{zhufangcheng1}),  one sees that for atoms aligned perpendicular to the boundary, the environment-induced interaction affects the entanglement dynamics for \emph{any} initial state, not only for a restricted class satisfying $\rho_{AS}(0),\,\rho_{SA}(0)\neq0$ as in free space~\cite{Chen}.  In the limit where both atoms are far from the boundary, the atom-boundary contribution becomes negligible and our results reduce to those of Ref.~\cite{Chen}, as expected.

In what follows, we focus on two representative initial states: the separable state \(|10\rangle\) (entanglement generation) and the maximally entangled antisymmetric state \(|A\rangle\) (entanglement degradation). For both cases, \(\rho_{GE}(\tau)=\rho_{EG}(\tau)=0\), so \(K_{2}(\tau)\le 0\) and
\bea\label{con}
C[\rho(\tau)]=\max\{0,K_{1}(\tau)\}.
\eea

\subsection{Entanglement generation for two-atom system with initial state $|10\rangle$}
We begin with the separable initial state \(|10\rangle\) and analyze entanglement generation. We first derive the early-time condition for entanglement creation and then discuss the subsequent evolution.

\subsubsection{Condition for entanglement generation}
First, we investigate entanglement generation at the neighborhood of the initial time. For a two-atom system prepared in the initial state $|10\rangle$,  one has $K_1(0)=0$.  Entanglement is generated at early times if $dK_{1}(\tau)/d\tau>0$  at $(\tau=0)$ . One finds
\bea\label{k1-28}
\frac{{K}_{1}(\tau)}{d\tau}\bigg|_{\tau=0}=4\sqrt{A_{2}^{2}+D^{2}}-4\sqrt{A_{1}^{2}-B_{1}^{2}},
\eea
which is independent of $\Delta$. Consequently,  the difference in the individual environment-induced energy shifts does not affect the \emph{early-time} entanglement-generation condition. Equation~\eqref{k1-28} further implies that, in this case,  the environment-induced interatomic interaction (the $D$ term) always enhances  entanglement generation, regardless of the interatomic separation or the distance of the two-atom system from the reflecting boundary.

\subsubsection{Time evolution of entanglement}

At general times, the closed-form solution is lengthy, but the early-time expansion is informative. For 
$\tau\to 0$, the leading term of the concurrence coefficient $K_{1}(\tau)$ is
\bea\label{EK1}
\nn K_{1}(\tau)&\approx&4\sqrt{B_{3}^{2}+D^{2}}\tau-4(3B_{1}+B_{2})\sqrt{B_{3}^{2}+D^{2}}\tau^{2}+\frac{2}{3}\sqrt{B_{3}^{2}+D^{2}}\tau^{3}\\
&&(28B_{1}^{2}+16B_{1}B_{2}+4B_{2}^{2}+16B_{3}^{2}-16D^{2}-\Delta^{2}).
\eea
 This expansion shows that the environment-induced interatomic interaction can enhance the generation of entanglement, while the environment-induced atom-boundary interaction tends to suppress it.   Since both mechanisms are present near the boundary, the net impact can be either constructive or destructive, in contrast to the boundary-free case~\cite{Chen}, where the induced interaction always assists entanglement generation.

\begin{figure}[htbp]
  \centering
  \includegraphics[width=0.32\textwidth]{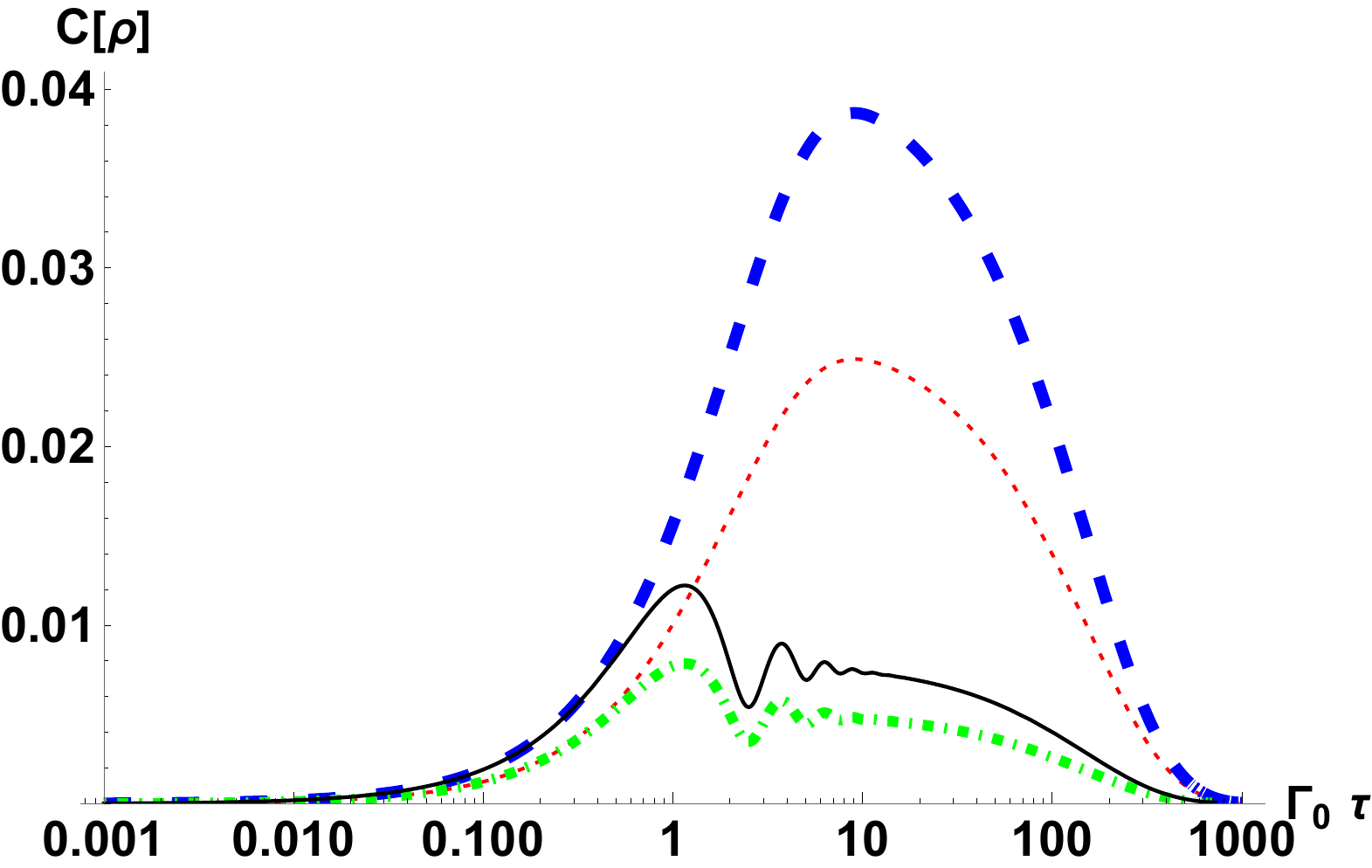}
  \includegraphics[width=0.32\textwidth]{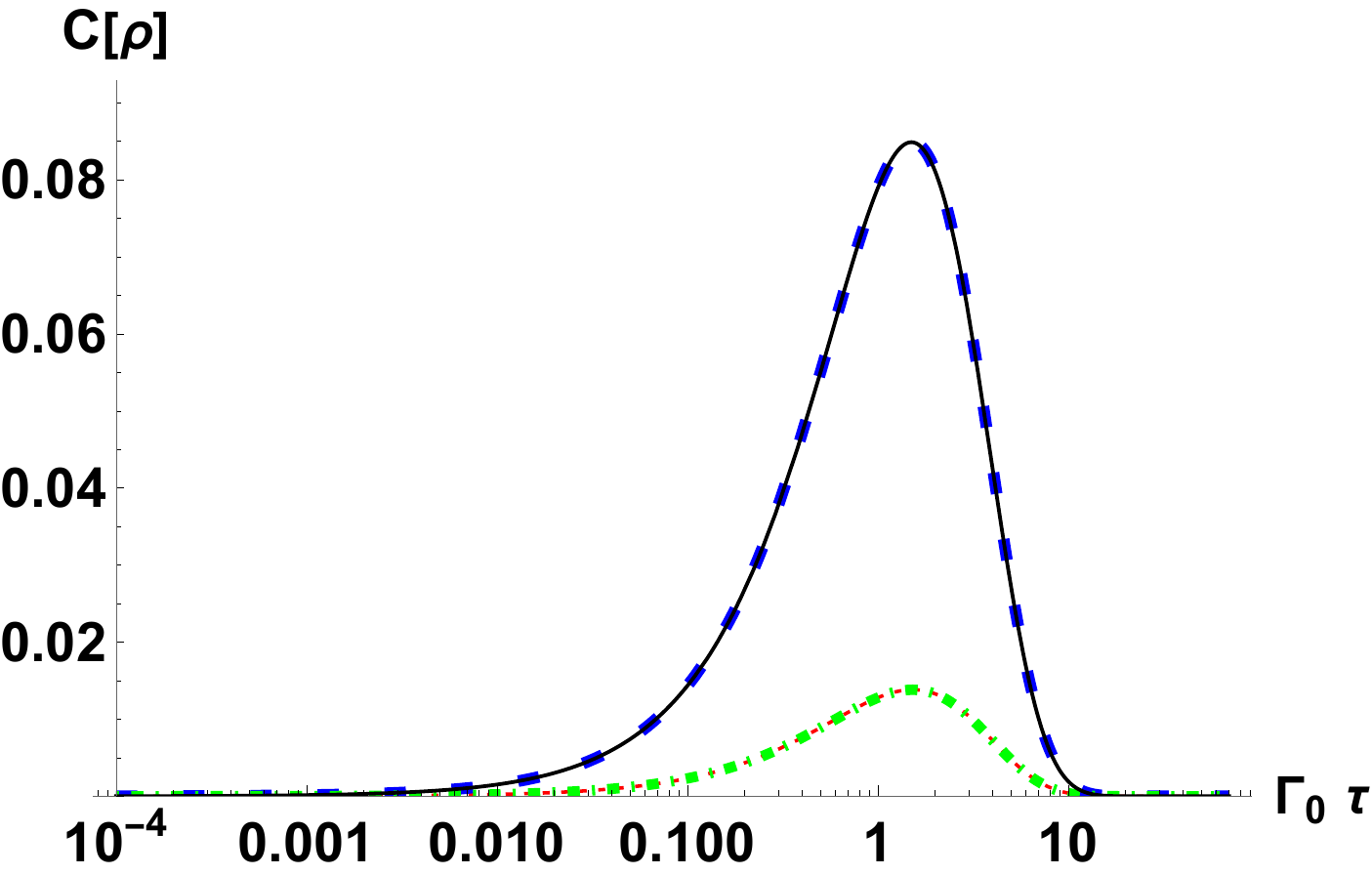}
  \includegraphics[width=0.32\textwidth]{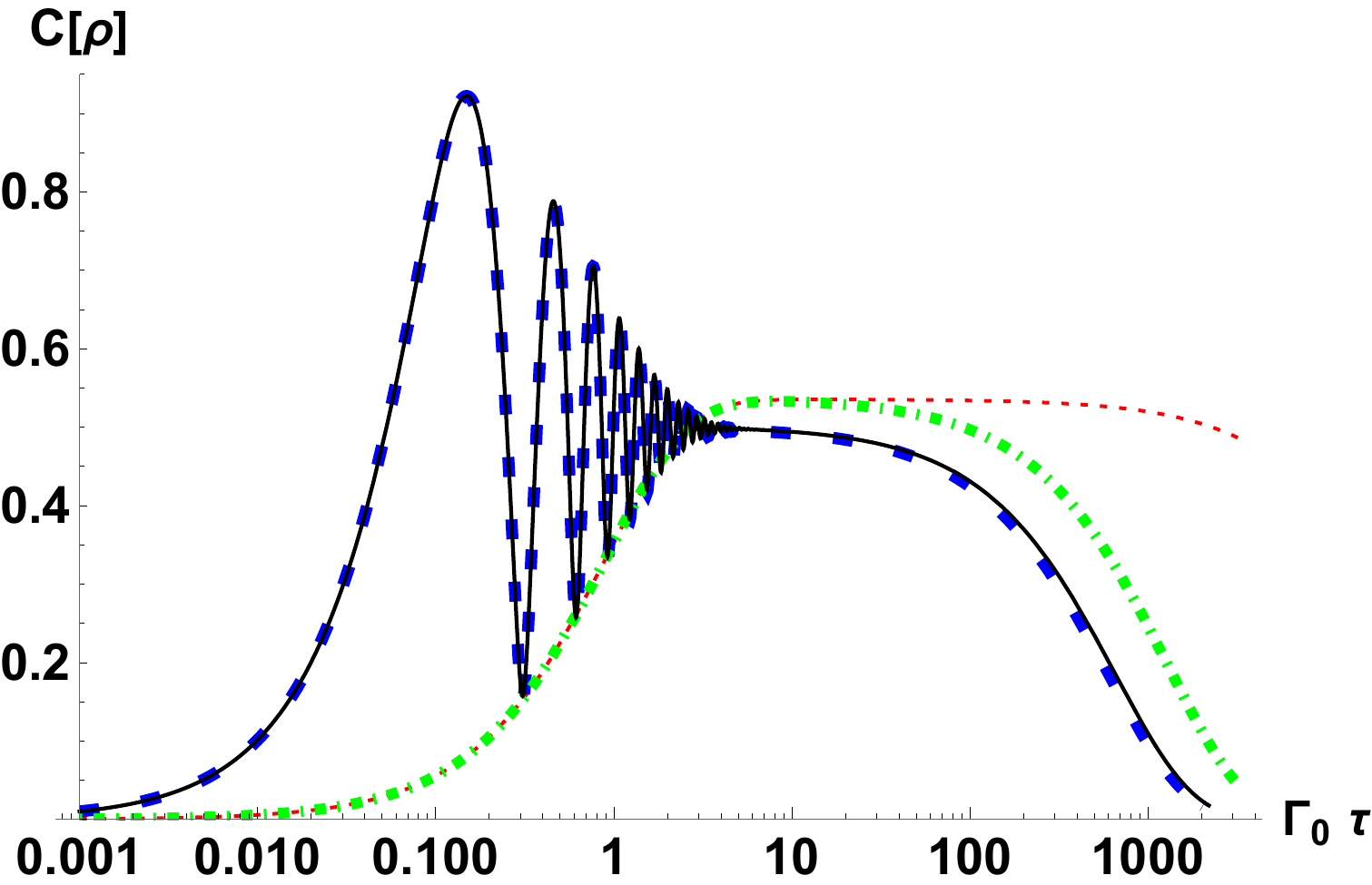}
  \caption{Time evolution of concurrence for atoms with (black solid line) and without (red dotted line) environment-induced interactions, with only the environment-induced atom-atom interaction~(blue dashed line), and with only the environment-induced atom-plate interaction~(green dot-dashed line). The parameters are $\omega y=\frac{1}{10},\,\,\omega L=10$ (left), $\omega y =1,\,\,\omega L=10 $ (middle), and $\omega y =1,\,\,\omega L=\frac{1}{10}$ (right).}\label{10}
\end{figure}
\begin{figure}[htbp]
  \centering
  \includegraphics[width=0.4\textwidth]{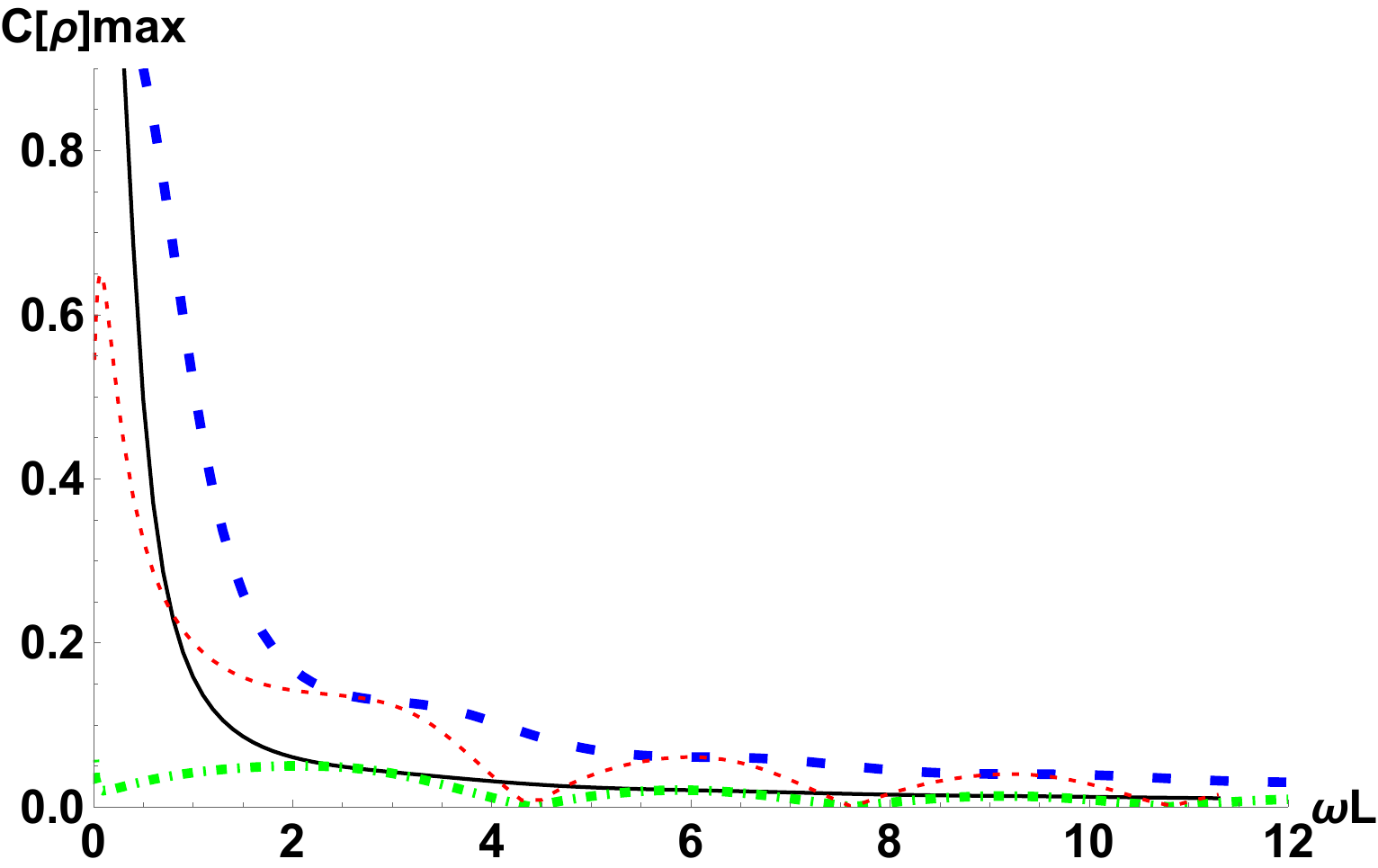}
  \includegraphics[width=0.4\textwidth]{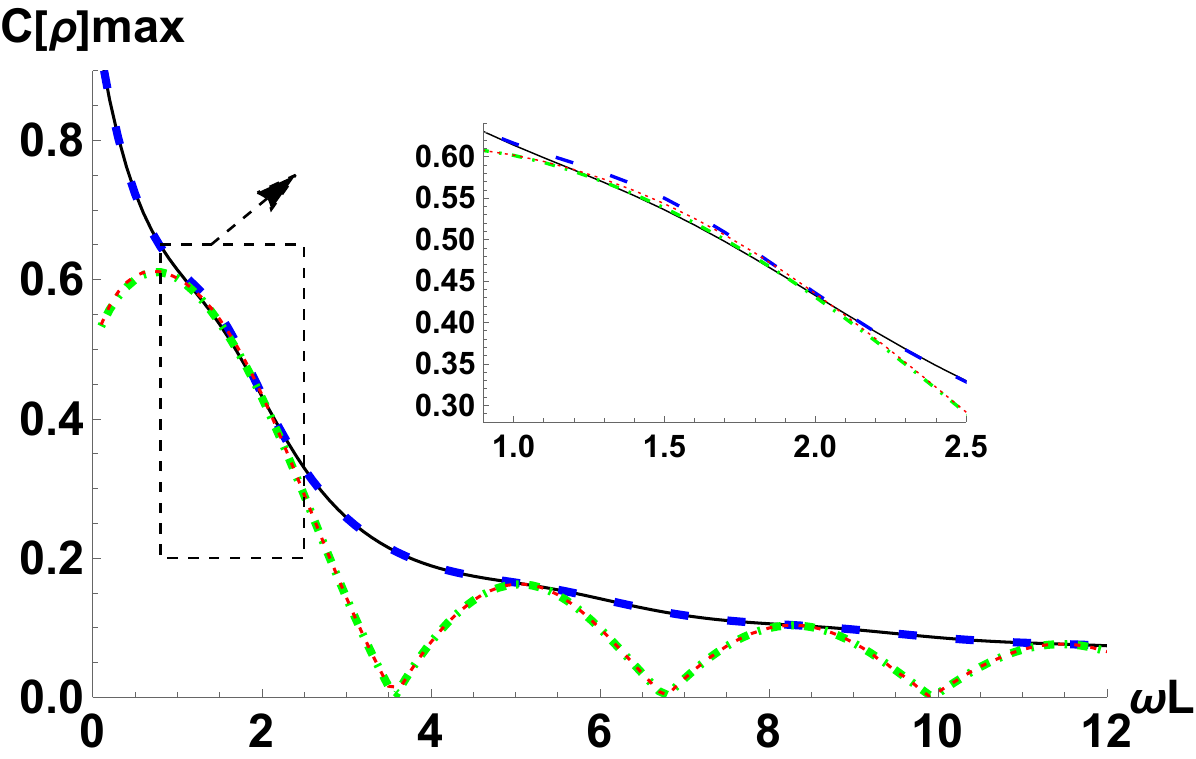}
  \caption{Maximum concurrence generated during evolution for atoms initially prepared in $|10\rangle$: with (black solid line) and without (red dotted line) environment-induced interactions, with only the environment-induced atom-atom interaction~(blue dashed line), and with only the environment-induced atom-plate interaction~(green dot-dashed line). The parameters are $\omega y = \frac{1}{10} $ (left) and $\omega y =1 $ (right). }\label{C2}
\end{figure}

In Fig.~\ref{10}, we show the time evolution of concurrence for a two-atom system initially in the state $|10\rangle$. 
In particular, we compare four cases: (i) atoms with both environment-induced interactions present ($D \neq 0$, $\Delta \neq 0$); (ii) without environment-induced interactions ($D = \Delta = 0$); (iii) with only the environment-induced atom-atom interaction ($D \neq 0$, $\Delta = 0$); and (iv) with only the environment-induced atom-boundary interaction ($D = 0$, $\Delta \neq 0$). 
When the distance between the nearer atom and plate is small compared with the transition wavelength, while the interatomic separation is larger than the transition wavelength [Fig.~\ref{10} (left)], the environment-induced interatomic interaction dominates at early times, thereby assisting entanglement generation. At later times, the environment-induced atom-plate interaction becomes dominant, suppressing entanglement generation.
When the distance between the nearer atom and plate is comparable to the transition wavelength and the interatomic separation is larger than the transition wavelength [Fig.~\ref{10} (middle)], the environment-induced atom-plate interaction has little influence on the evolution of entanglement. In this case, the environment-induced interatomic interaction plays a dominant role, assisting the entanglement generation.
When the distance between the nearer atom and plate is comparable to the transition wavelength and the interatomic separation is small compared with the transition wavelength [Fig.~\ref{10} (right)], the environment-induced interatomic interaction dominates at early times, enhancing entanglement generation. At later times, however, both the environment-induced interatomic interaction and the environment-induced atom-plate interaction affect the evolution of entanglement. Their combined effect suppresses entanglement generation, leading to a shorter survival time for the entanglement.

Next, we study the maximum concurrence generated during the evolution. As shown in Fig.~\ref{C2} (left), when the separation between the nearer atom and the plate is small compared to the transition wavelength, the maximum concurrence generated is always larger for two-atom systems with a small interatomic separation compared to the transition wavelength when the environment-induced interaction is considered. In this regime, the environment-induced interatomic interaction plays a dominant role. However, when the interatomic separation is comparable to or larger than the transition wavelength, the maximum concurrence can be either smaller or larger, depending on whether the environment-induced interatomic interaction or the environment-induced atom-plate interaction is dominant.
When the separation between the nearer atom and the plate is comparable to the transition wavelength [Fig.~\ref{C2} (right)], the maximum concurrence generated when the environment-induced interactions are considered can only be smaller in a small region around $\omega L \sim 1.5$ [Fig.~\ref{C2} (right), inset]. In this case, the environment-induced atom-plate interaction plays a dominant role.

\subsection{Entanglement degradation for two-atom system with initial state $|A\rangle$}

We next consider the maximally entangled antisymmetric state $|A\rangle$. We note that, under the dynamics considered here, the symmetric state $|S\rangle$ exhibits entanglement decay behavior qualitatively similar to that of $|A\rangle$ and therefore does not provide additional
independent insight into the main physical mechanisms we aim to highlight. For this reason, we take $|A\rangle$ as a representative example.

At early times ($\tau\to 0$), the leading term of the concurrence coefficient $K_{1}(\tau)$ can be expressed as
\bea\label{AK1}
K_{1}(\tau)&\approx&1-2\tau(B_{1}+B_{2}-2B_{3})+2\tau^{2}(B_{1}+B_{2}-2B_{3})^{2}-\\
\nn&& \frac{4}{3}\tau^{3}[B_{1}^{3}+B_{2}^{3}+B_{1}^{2}(3B_{2}-8B_{3})-8B_{2}^{2}B_{3}+12B_{2}B_{3}^{2}-8B_{3}^{3}\\
\nn&& +B_{1}(3B_{2}^{2}-8B_{2}B_{3}+12B_{3}B_{2})-4B_{2} D \Delta+4B_{1} D \Delta+2B_{3}\Delta^{2}].
\eea
From Eq. (\ref{AK1}), we find that in the regime where the nearer atom is close to the boundary, namely $y \ll 1/\omega$ and $y \ll L$, such that $|\Delta|\gg |D|$, the concurrence is reduced and the entanglement decays more rapidly when the environment-induced interactions are considered. When the interatomic separation is much smaller than the transition wavelength, i.e.,  $L \ll 1/\omega$, the environment-induced interatomic interaction $D$ becomes large, and one has $B_{1}>0, B_{2}>0, D>0$, and $\Delta<0$. When we choose an appropriate distance $y$ between the atoms and the boundary such that $B_{1}<B_{2}$, the concurrence is again reduced and the entanglement decays faster in the presence of environment-induced interactions.

\begin{figure}[htbp]
  \centering
  \includegraphics[width=0.32\textwidth]{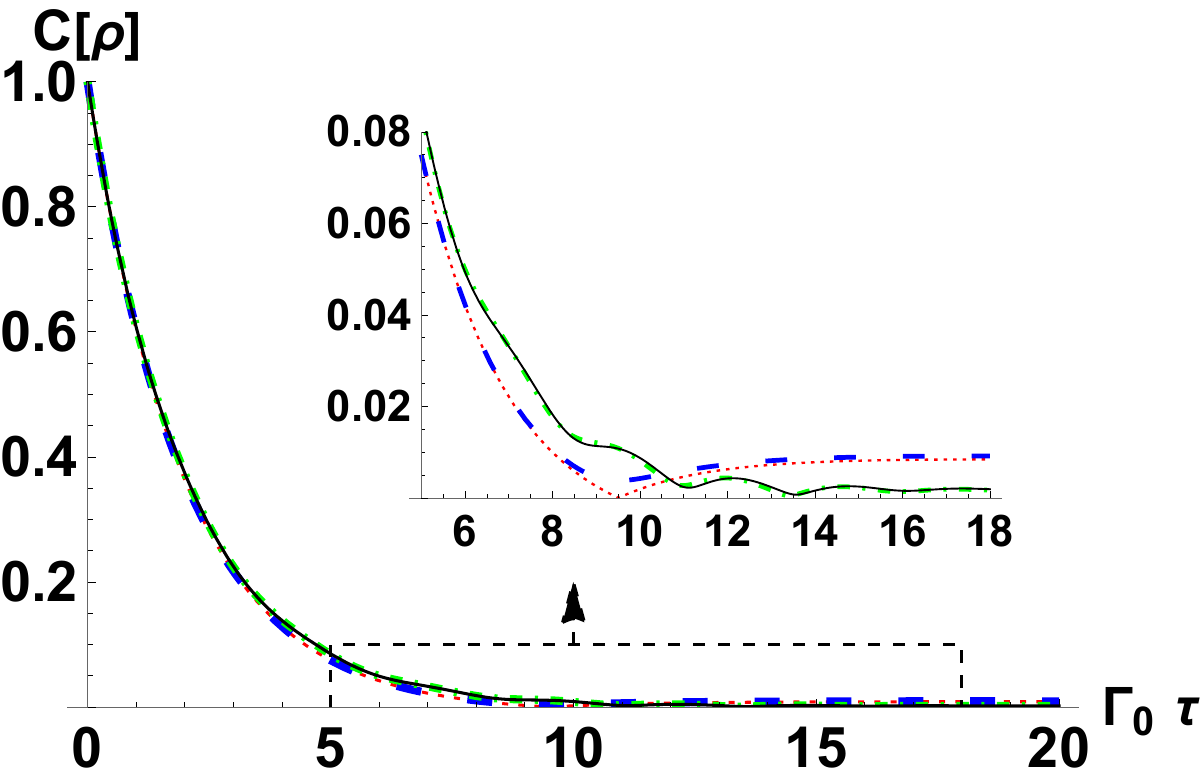}
  \includegraphics[width=0.32\textwidth]{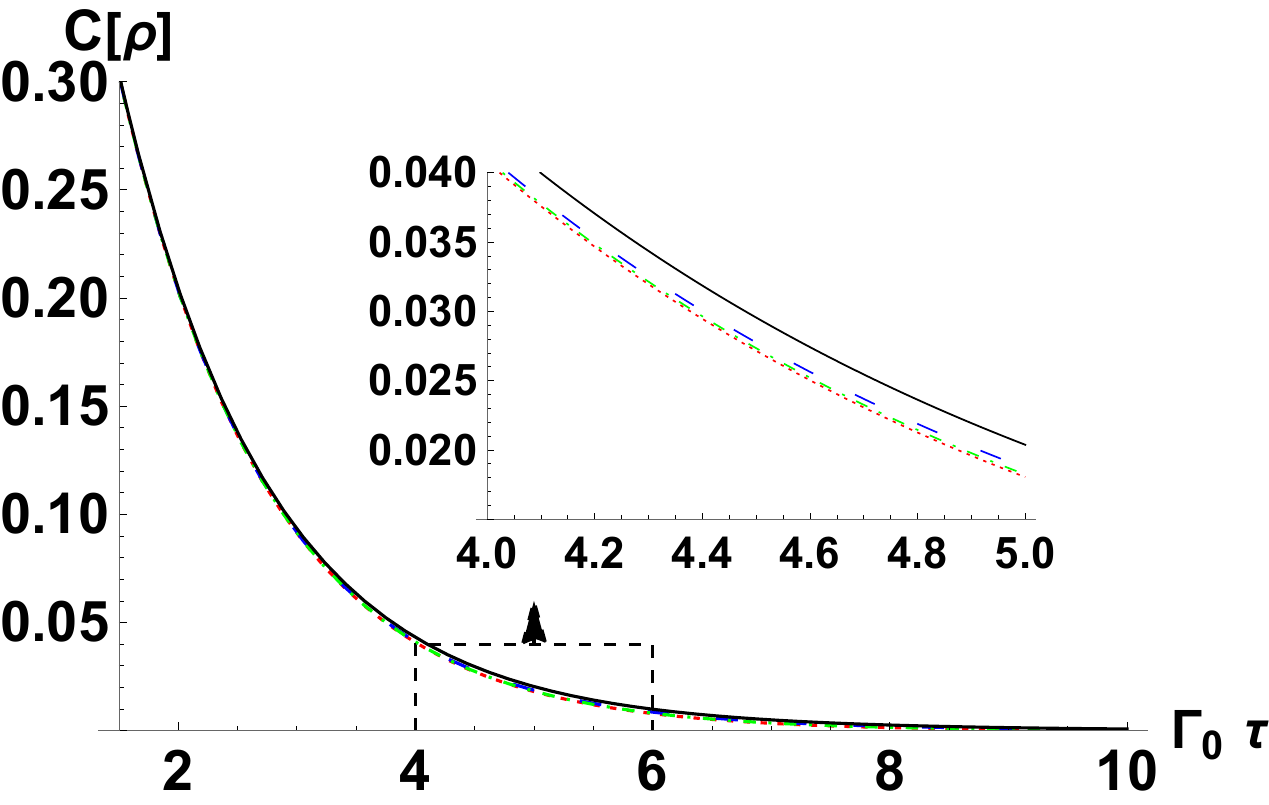}
  \includegraphics[width=0.32\textwidth]{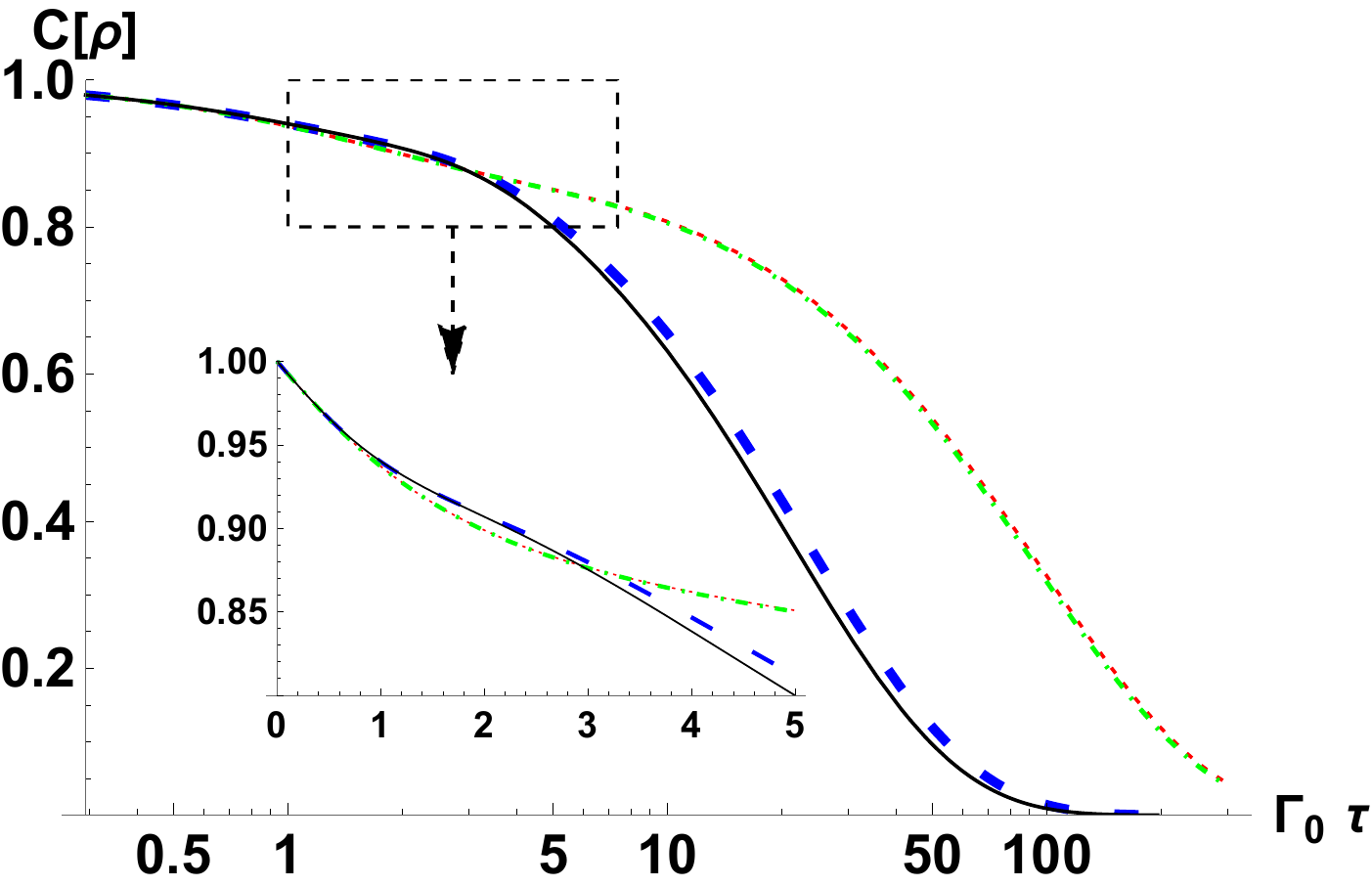}
  \caption{Time evolution of concurrence  for atoms with (black solid line) and without (red dotted line) environment-induced interactions, with only the environment-induced atom-atom interaction~(blue dashed line), and with only the environment-induced atom-plate interaction~(green dot-dashed line). The parameters are $\omega y=\frac{1}{10},\,\,\omega L=19$ (left), $\omega y =1,\,\,\omega L=19 $ (middle), and $\omega y =1,\,\,\omega L=1$ (right).}\label{A}
\end{figure}

In Fig.~\ref{A}, we present the time evolution of entanglement for the two-atom system. When the distance between the nearer atom and the plate is small compared with the transition wavelength, while the interatomic separation is larger than the transition wavelength [Fig.~\ref{A} (left)], the environment-induced interactions have little influence on the entanglement dynamics at early times. At later times, however, the environment-induced atom-plate interaction becomes dominant. In this regime, the environment-induced interactions initially slow down the decay of entanglement and subsequently accelerate it, resulting in a shorter entanglement survival time.
When the distance between the nearer atom and the plate is comparable to the transition wavelength, and the interatomic separation is larger than the transition wavelength [Fig.~\ref{A} (middle)], the environment-induced atom-plate interaction has little effect on the entanglement evolution. Instead, the environment-induced interatomic interaction dominates, resulting in a slower decay of entanglement and, consequently, a longer entanglement lifetime. 
When both the distance between the nearer atom and the plate and the interatomic separation are comparable to the transition wavelength [Fig.~\ref{A} (right)], the environment-induced interatomic interaction dominates at early times, again leading to a slower decay of entanglement. At later times, both the environment-induced interatomic interaction and the environment-induced atom-plate interaction significantly influence the entanglement dynamics. Their combined effect accelerates entanglement decay, thereby shortening the entanglement survival time.

\section{Discussion}

In this work, we adopt a Dirichlet boundary condition for the scalar field as an idealized analog of a perfectly reflecting surface. 
This model becomes unphysical arbitrarily close to the boundary, where the atom-boundary interaction contribution diverges [see Eqs.~\eqref{d}-\eqref{f2}]. This is a standard limitation of idealized boundary conditions and should not be interpreted as a feature of realistic material systems.

In realistic physical systems, however, this divergence is  regularized for several reasons. First, an atom cannot be treated as a pointlike object at distances comparable to its finite size and therefore cannot be placed arbitrarily close to the boundary. Second, real boundaries are not perfectly reflecting at all frequencies. For example, a conducting plate reflects electromagnetic waves efficiently only below its plasma frequency and becomes increasingly transparent at higher frequencies,  where the idealized Dirichlet condition ceases to be valid. Third, in the present model, both the boundary and the atomic positions are assumed to be fixed. Allowing for their position fluctuations would also regularize the energy divergence (see, e.g., Refs.~\cite{Ford97,Ford98}). 

At the same time,  we would like to emphasize that the central observable studied here, namely the concurrence, remains finite and well defined even in the near-boundary limit.  In particular,  when the nearer atom approaches the boundary ($y \to 0$), one finds $D \to 0$, $\Delta \to \infty$, and the dissipation coefficient $B_3\to0$. In this limit, for an initial separable state such as $|10\rangle$, the concurrence remains zero throughout the evolution, so no entanglement is generated, and the divergence in $\Delta$ has no physical impact on the entanglement. For an initial entangled state such as the antisymmetric state $|A\rangle$, the concurrence takes the form  $K_1 (\tau) = \sqrt{e^{-4 (B_1+B_2)\tau} \cos(4 \Delta \tau)}$, so the large value of $\Delta$ produces only rapid bounded oscillations rather than any divergence. 
Thus,  the idealized divergence of the atom-boundary contribution does not lead to any divergence of the entanglement measure itself. Its effect is instead to suppress entanglement generation or to induce bounded high-frequency oscillations.

\section{Conclusion}
We have investigated, within the open-quantum-system framework, how environment-induced interactions modify the entanglement dynamics of two atoms weakly coupled to a fluctuating scalar field in the Minkowski vacuum, with their positions fixed near a perfectly reflecting boundary. We have focused on a vertical atomic configuration relative to the boundary, where position-dependent Lamb shifts introduce a nontrivial atom-boundary contribution in addition to the field-mediated atom-atom interaction.

A key qualitative finding is that, near the boundary, environment-induced interactions affect the entanglement dynamics for \emph{any} initial two-atom state. This contrasts with the free-space case~\cite{Chen}, where the induced interaction influences entanglement generation only for a restricted class of initial states. As expected, in the far-field limit where atom-boundary effects become negligible, our results reduce to those in free space.

For atoms initially prepared in a separable state with one atom in the ground state and the other in the excited state, the atom-boundary contribution does not alter the early-time condition for entanglement generation, whereas the induced atom-atom interaction enhances the initial entanglement growth rate. At later times, however, the two mechanisms compete: depending on which contribution dominates, the maximum concurrence and the entanglement lifetime may either increase (when the induced atom-atom interaction dominates) or decrease (when atom-boundary effects dominate). This behavior is qualitatively different from the boundary-free case, where the induced interaction assists entanglement generation without exception.

For atoms initially prepared in the antisymmetric maximally entangled state $|A\rangle$, the environment-induced interactions have little influence on the entanglement dynamics at early times, but become important at intermediate and later times, reflecting again the competition between atom-boundary and induced atom-atom effects. When the nearer atom lies very close to the boundary, i.e., when the atom-boundary separation is much smaller than the transition wavelength while the interatomic separation is large, the environment-induced interactions first slow down entanglement decay and then accelerate it, ultimately shortening the entanglement survival time.  A similar ``suppression-enhancement'' sequence occurs when both the atom-boundary separation and the interatomic separation are comparable to the transition wavelength, also resulting in a shorter entanglement lifetime.  In contrast, when the atom-boundary separation is comparable to the transition wavelength and the interatomic separation is large, the environment-induced interatomic interaction dominates, yielding a slower decay of entanglement and hence a longer entanglement lifetime.

\begin{acknowledgments}
This work was supported in part by the NSFC under Grants No. 12075084 and 12575051, the innovative research group of Hunan Province under Grant No. 2024JJ1006, and the Doctoral Research Initiation Project of China West Normal University under Grant No. 23KE026.
\end{acknowledgments}

\end{document}